\documentclass[aps,prd,twocolumn,superscriptaddress,showpacs,amsfonts,amssymb,amsmath]{revtex4}
\usepackage[dvips]{graphicx}
\DeclareGraphicsExtensions{.ps,.eps,.pdf}
 \DeclareGraphicsExtensions{.jpg,.png}
\usepackage{color}
\usepackage{ulem}
\usepackage[colorlinks=true,linkcolor=blue,citecolor=blue,urlcolor=blue]{hyperref}
\usepackage{tabularx}
\usepackage{multirow}

\begin{document}

\title{
Helical bunching and symmetry lowering inducing multiferroicity in Fe langasites}

\author{L. Chaix}
\affiliation{Institut Laue-Langevin, 6 rue Jules Horowitz, 38042 Grenoble, France}
\affiliation{Institut N\'eel, CNRS, 38042 Grenoble, France}
\affiliation{Universit\'e Grenoble Alpes, 38042 Grenoble, France}
\affiliation{Stanford Institute for Materials and Energy Sciences, SLAC National Accelerator Laboratory, Menlo Park, California 94025, USA}
\author{R. Ballou}
\affiliation{Institut N\'eel, CNRS, 38042 Grenoble, France}
\affiliation{Universit\'e Grenoble Alpes, 38042 Grenoble, France}
\author{A. Cano}
\affiliation{CNRS, Univ. Bordeaux, ICMCB, UPR 9048, F-33600 Pessac, France}
\author{S. Petit}
\affiliation{CEA, Centre de Saclay, /DSM/IRAMIS/ Laboratoire L\'eon Brillouin, 91191 Gif-sur-Yvette,France}
\author{S. de Brion}
\affiliation{Institut N\'eel, CNRS, 38042 Grenoble, France}
\affiliation{Universit\'e Grenoble Alpes, 38042 Grenoble, France}
\author{J. Ollivier}
\affiliation{Institut Laue-Langevin, 6 rue Jules Horowitz, 38042 Grenoble, France}
\author{L.-P. Regnault}
\affiliation{SPSMS-MDN, INAC, 38054 Grenoble, France}
\affiliation{Universit\'e Grenoble Alpes, 38042 Grenoble, France}
\author{E. Ressouche}
\affiliation{SPSMS-MDN, INAC, 38054 Grenoble, France}
\affiliation{Universit\'e Grenoble Alpes, 38042 Grenoble, France}
\author{E. Constable}
\affiliation{Institut N\'eel, CNRS, 38042 Grenoble, France}
\affiliation{Universit\'e Grenoble Alpes, 38042 Grenoble, France}
\author{C. V. Colin}
\affiliation{Institut N\'eel, CNRS, 38042 Grenoble, France}
\affiliation{Universit\'e Grenoble Alpes, 38042 Grenoble, France}
\author{A. Zorko}
\affiliation{Jo\v{z}ef Stefan Institute, Jamova 39, 1000 Ljubljana, Slovenia}
\author{V. Scagnoli}
\affiliation{Laboratory for Mesoscopic Systems, Department of Materials, ETH Zurich, 8093 Zurich, Switzerland}
\affiliation{Paul Scherrer Institute, 5232 Villigen PSI, Switzerland}
\author{J. Balay}
\affiliation{Institut N\'eel, CNRS, 38042 Grenoble, France}
\affiliation{Universit\'e Grenoble Alpes, 38042 Grenoble, France}
\author{P. Lejay}
\affiliation{Institut N\'eel, CNRS, 38042 Grenoble, France}
\affiliation{Universit\'e Grenoble Alpes, 38042 Grenoble, France}
\author{V. Simonet}
\affiliation{Institut N\'eel, CNRS, 38042 Grenoble, France}
\affiliation{Universit\'e Grenoble Alpes, 38042 Grenoble, France}

\date{\today}

\begin{abstract}
The chiral Fe-based langasites represent model systems of triangle-based frustrated magnets with a strong potential for multiferroicity. We report neutron scattering measurements for the multichiral Ba$_3M$Fe$_3$Si$_2$O$_{14}$ ($M$ = Nb, Ta) langasites revealing new important features of the magnetic order of these systems: the bunching of the helical modulation along the $c$-axis and the in-plane distortion of the 120$^{\circ}$ Fe-spin arrangement. We discuss these subtle features in terms of the microscopic spin Hamiltonian, and provide the link to the magnetically-induced electric polarization observed in these systems. Thus, our findings put the multiferroicity of this attractive family of materials on solid ground.

\end{abstract}


\maketitle




The quest for novel magnetic orders that promote multiferroicity and strong magnetoelectric coupling drives a considerable amount of current research
(see e.g. \cite{TOK15,DON15,KHO15,BOS16} for recent reviews). In this respect, Fe-based langasites are propitious materials displaying an original ``multi-chiral'' magnetic order in which a 120$^{\circ}$ Fe-spin arrangement undergoes an additional helical modulation \cite{MIL98,MAR08,SIM12,YU09}. In fact, the magnetoelectric effect in these systems has been demonstrated at both static \cite{MAR10} and dynamical levels \cite{CHA13}. In addition, Zhou {\it et al.} \cite{ZHO09} have reported the emergence of ferroelectricity at the onset of  magnetic order. This ferroelectricity has subsequently been confirmed by Lee et al. \cite{LEE14}, who have also demonstrated the magnetic-field manipulation of the electric polarization. However, these works report two different ferroelectric axis at zero-field and, above all, leave the precise mechanism promoting such a ferroelectricity unidentified. 

Magnetically-induced ferroelectricity in many multiferroic cycloidal magnets is due to the spin-orbit coupling (SOC) via the so-called inverse Dzyaloshinskii-Moryia (DM) mechanism \cite{Sergienko06,Katsura05,MOS06}. However, this mechanism is ineffective for the reported helical magnetic structure of the langasites \cite{WHI13,CAO14}. An electric polarization observed in magnetic helices rather than cycloids has been theorized \cite{ARI07}, as well as observed experimentally in RbFe(MoO$_4$)$_2$ for instance \cite{KEN07,WHI13,CAO14}. However, the invoked mechanisms still rely on the SOC and appear inapplicable in the langasite case. Bulaevskii {\it et al.} \cite{BUL08,KHO10,KHO15} have discussed another fundamental mechanism that can generically operate in triangle-based frustrated magnets. The proposed mechanism is independent of SOC as it simply relies, in the language of Mott insulators, on high-order hopping-term contributions that enable the redistribution of the electronic charge. This purely electronic mechanism can be supplemented by a contribution from lattice degrees of freedom \cite{KHO15}. In any case, it requires a distortion of the perfect 120$^\circ$-spin structure and, despite its generality, has not been verified experimentally so far.

In this paper, we carefully analyze neutron scattering experiments to revise the magnetic structure of the Ba$_3 M$Fe$_3$Si$_2$O$_{14}$ ($M=$ Nb, Ta) langasites. We discover additional features indicating  the bunching of the helicoidal modulation and significant deviations from the 120$^\circ$-spin structure in these systems. In order to explain these features, we propose an extended spin Hamiltonian incorporating different magnetic anisotropies. These extra terms prove to be crucial for understanding the magnetically-induced ferroelectricity in the Fe langasites. Furthermore, we find that this ferroelectricity emerges from a magnetoelectric coupling that is equivalent to the Bulaevskii {\it et al.} proposal at the phenomenological level. 

\begin{figure}
\resizebox{8.6cm}{!}{\includegraphics{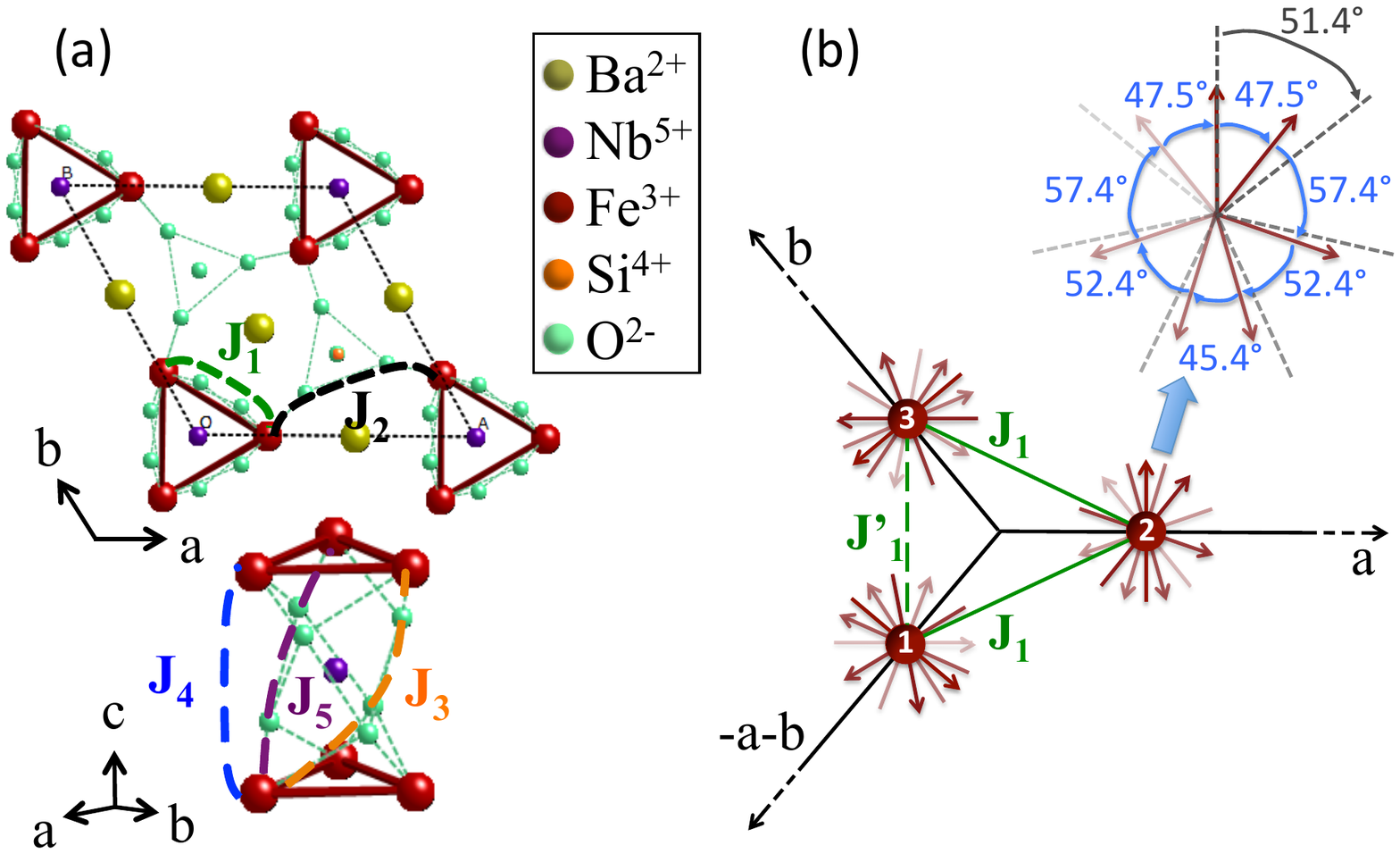}}
\caption{(Color online) (a) Ba$_3$NbFe$_3$Si$_2$O$_{14}$ crystallographic structure in the $ab$-plane and along the $c$-axis, with five magnetic exchange paths $J_{1}$ to $J_{5}$. (b) Schematic representation of the revisited magnetic structure projected along the $c$-axis. The 3-fold axis is broken leading to different $J_1$ interactions (case of an isosceles triangle represented). The bunching of the helices in the Nb compound is reported through the successive angles between consecutive magnetic moments along $c$, deviating from the regular angle $2\pi\tau$=51.4$^{\circ}$.}
\label{fig1c}
\end{figure}

The Fe langasites crystallize in the non-polar non-centrosymmetric space group P321, in which the magnetic Fe$^{3+}$ ions ($S=5/2$) form a triangular lattice of triangles in the $ab$-plane (see Fig. \ref{fig1c}). The magnetic order of these systems, observed below the N\'eel temperature T$_N$, has been rationalized within the spin Hamiltonian \cite{MAR08,LOI11,ZOR11,JEN11,SIM12}:
\begin{equation}
\mathcal{H}=  \sum _{(ij)}J_{ij} {{\bf S}_{i} \cdot {\bf S}_{j}} + \sum _{(ij)_{\triangle}}
D ({{\bf S}_{i} \times {\bf S}_{j}})_z.
\label{eq1}
\end{equation}

Here the first term takes into account the exchange interactions, where the sum runs over the spin pairs connected by different paths shown in Fig. \ref{fig1c} (a). The nearest- and next-nearest-neighbor interactions $J_1$ and $J_2$ in the $ab$-plane are both antiferromagnetic. This generates magnetic frustration within the Fe triangles and causes the 120$^\circ$ spin structure. Among the three antiferromagnetic exchange couplings between adjacent $ab$-planes, the strongest one is $J_3$ or $J_5$ depending on the structural chirality. These couplings generate the helical modulation of the 120$^\circ$ structure along the $c$-axis. The propagation vector of this modulation depends on the balance between the inter-plane interactions $J_3$, $J_4$ and $J_5$. The second term in Eq. \eqref{eq1}, where the sum runs over intra-triangle spins, corresponds to the additional DM interactions with the DM vector ($D$) along the $c$-axis, which is allowed by symmetry. This DM term restricts the spins to the $ab$-plane and favors one intra-triangle chirality as observed experimentally \cite{SM,LOI11,ZOR11}. Alternatively, a weak easy-axis single-ion anisotropy can have similar effect \cite{STO11}.

\begin{figure*}
\resizebox{18cm}{!}{\includegraphics{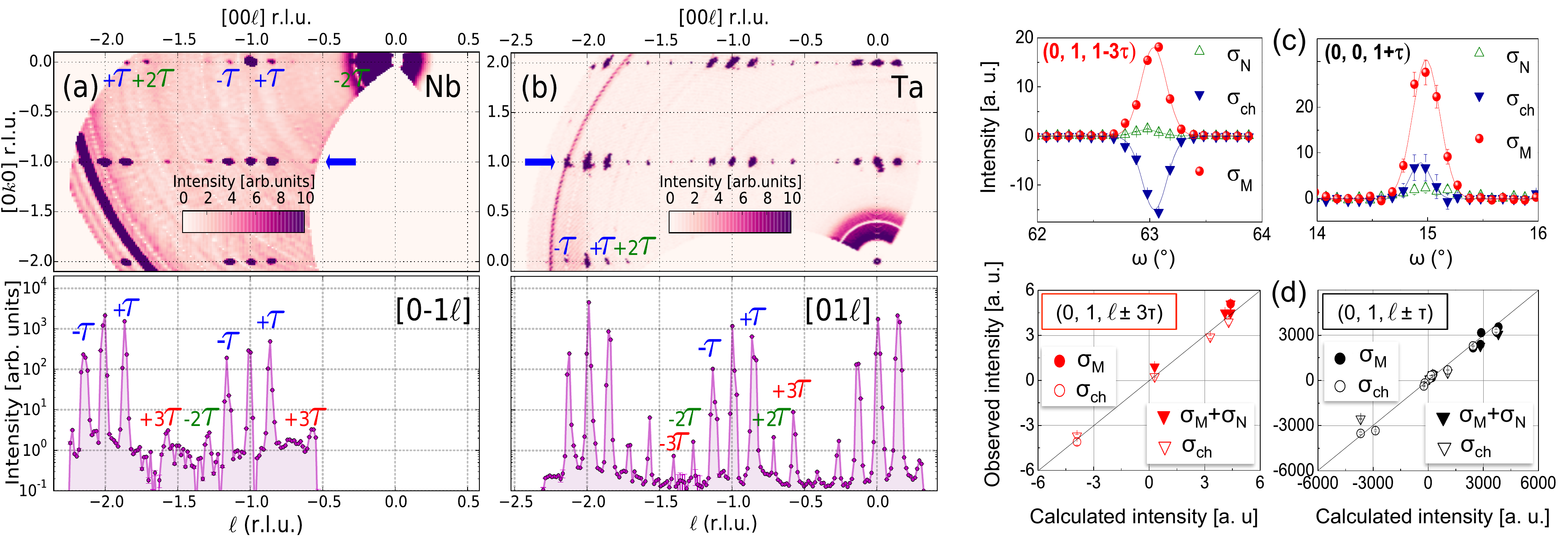}}
\caption{(color online) Zero energy cut (integrated from -0.1 to 0.1 meV) from IN5 measurements at 1.5 K, on the Nb (a) and Ta (b) compounds. The blue arrows on the energy maps indicated the [0 $\pm$1 $\ell$]* directions of the $Q$-1D cuts shown below in vertical log scale. (c)-(d) summarize the measurements performed on the Nb compound at 2 K on IN22 with polarized neutrons. (c) shows the magnetic (red), nuclear (green) and chiral (blue) elastic contributions from scans obtained by rotating the crystal ($\omega$-scans) with the H-H configuration. The solid lines are gaussian fits. The magnetic and chiral contributions measured in the H-H (circles) and G-H (triangles) configurations for the main (0, 1, $\ell \pm \tau$) and extra (0, 1, $\ell \pm 3\tau$) satellites, versus the calculated contributions using the model given in the caption of Fig. \ref{fig2}, are displayed in (d). Note that the G-H configuration cannot separate the magnetic from nuclear scattering, but we checked that $\sigma_M+\sigma_N\approx\sigma_M$ for these satellites.}
\label{fig1}
\end{figure*}


We have revised this structure by carrying out single-crystal neutron scattering measurements on two members of the Fe langasite family with Nb or Ta on the $1a$ Wyckoff site. The two single-crystals, grown by floating-zone method in an image furnace \cite{MAR10}, display opposite structural chiralities \cite{SM}. The experiments were done on two spectrometers installed at the Institut Laue Langevin: IN5 time-of-flight spectrometer using unpolarized neutrons, and CRG-CEA IN22 triple-axis spectrometer, equipped with CRYOPAD, using polarized neutrons and allowing longitudinal polarization analysis. The dynamical structure factors were measured on IN5 at 1.5 K, for both crystals in rotation about the $a$ zone axis, with an incident wavelength of 4 \AA. The Nb compound was also measured on IN22 at 2 K with a final wavevector fixed to 2.66 \AA$^{-1}$ and at zero energy transfer in order to get a quantitative determination of the elastic structure factors of the magnetic satellites. The crystal was oriented with the $a$ zone axis vertical so as to probe the ($b$*, $c$*) scattering plane. Two types of measurements were carried out using different monochromator-analyser configurations: Heusler-Heusler (H-H) and Graphite-Heusler (G-H). The former allows us to discriminate the magnetic and chiral contributions from the nuclear one, whereas the latter increases the incident neutron flux and allows us to determine the chiral scattering from an unpolarized beam \cite{SIM12}, hence with a larger flux. 

The intensity maps at zero energy cut measured on IN5 for the Nb and Ta compounds are shown in Figs. \ref{fig1}(a) and (b) respectively. The N\' eel temperature of these compounds are nearly identical (27 K for Nb and 28 K for Ta), while the propagation vectors are $\vec \tau= (0,0,0.1429)$ and $(0, 0, 0.1385)$ respectively \cite{MAR10,LYU11}. The latter is obtained from the position of the magnetic satellites at $Q = |\vec{H} \pm \vec{\tau}|$ in Figs.~\ref{fig1}. In addition to these first-order satellites, we clearly observe higher-order satellites at $Q= \mid \vec{H} \pm 2\vec{\tau} \mid$ and $Q = \mid \vec{H} \pm 3\vec{\tau} \mid$ in both compounds suggesting that the helical modulation is distorted. 

Polarized neutron measurements with longitudinal polarization analysis on the Nb compound allowed us to extract the nuclear ($\sigma_N$), the magnetic ($\sigma_M$) and the chiral ($\sigma_{ch}$) scattering \cite{SM}, hence to identify the nuclear and/or magnetic nature of these extra peaks. This analysis reveals that the second-order satellites are mainly of nuclear origin. This is usually attributed to magnetoelastic effects and will not be further discussed. In contrast, the magnetic contribution is dominant in the third-order satellites [see Fig.~\ref{fig1}(c)], which confirms the deformation of the magnetic helices. 

\begin{figure}
\resizebox{8.6cm}{!}{\includegraphics{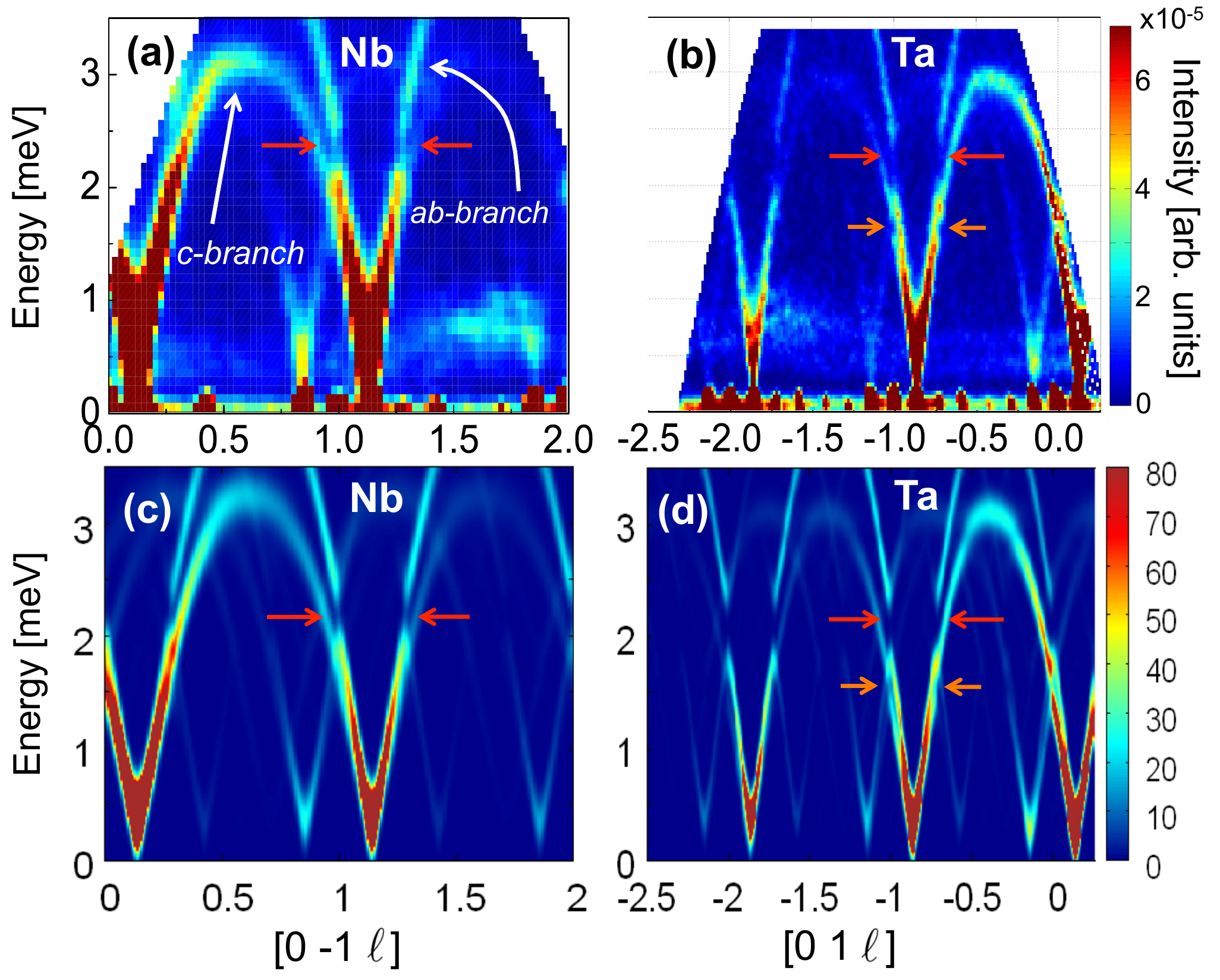}}
\caption{(Color online) Spin-waves spectra of Nb (a) and Ta (b) measured along the [0 -1 $\ell$]* and [0 1 $\ell$]* directions respectively at 1.5 K on IN5. Below: Spin-waves calculations with the following parameters for the Nb(c)/Ta(d): $J_{1} = 0.85/0.85\pm0.1$ meV, $J_{2} = 0.24/0.24\pm0.05$ meV, $J_{3} = 0.053/0.22\pm0.03$ meV, $J_{4} = 0.017/0.014\pm0.05$ meV and $J_{5} = 0.24/0.053\pm0.05$ meV, $D$= -0.028/-0.028 $\pm 0.005$ meV, easy-plane anisotropy $K=0.045/0.053\pm 0.005$ meV. The red/orange arrows show the spectral weight extinctions.}
\label{fig2}
\end{figure}

This deformation allows further explanation of the subtle details in the spin-wave spectrum. As we can see in Fig. \ref{fig2}, the spectra of the Nb and Ta compounds display two branches emerging from the magnetic satellites \cite{LOI11}. The high energy branch corresponds to spin components in the $ab$-plane while the low-energy branch is associated to spin components along the $c$-axis. The main features of this spectrum were initially captured by the model of Eq. 1 using the standard Holstein-Primakov formalism in the linear approximation (see Fig. 4 of ref. \onlinecite{LOI11}). However, this simple model fails to reproduce a spectral weight extinction observed in the $ab$-branch around 2.3 meV in both compounds, and a tiny extinction around 1.7 meV in the $c$-branch, visible only in the Ta compound (red and orange arrows respectively in Fig.~\ref{fig2}) \cite{SM}.

In order to reproduce the deformation of the helices leading to all these features, we have considered a straightforward extension of Eq. 1. We included extra single-ion anisotropy terms with respect to a local 2-fold $z_{loc}$ axes, {\it i.e.} along $a$, $b$ and $-a-b$ for the three ions of a triangle respectively:
\begin{align}
H_{sia}=  K \sum_{i} S_{z_{loc}, i}^2.
\label{Hsia}\end{align}
We first determine the magnetic ground state of the extended Hamiltonian by means of a real-space mean-field energy minimization. Then we compute the corresponding magnetic and chiral elastic scatterings for the first and third order magnetic satellites, and finally we compute the corresponding spin-wave spectrum \cite{SM}. 

\begin{figure}
\resizebox{8.6cm}{!}{\includegraphics{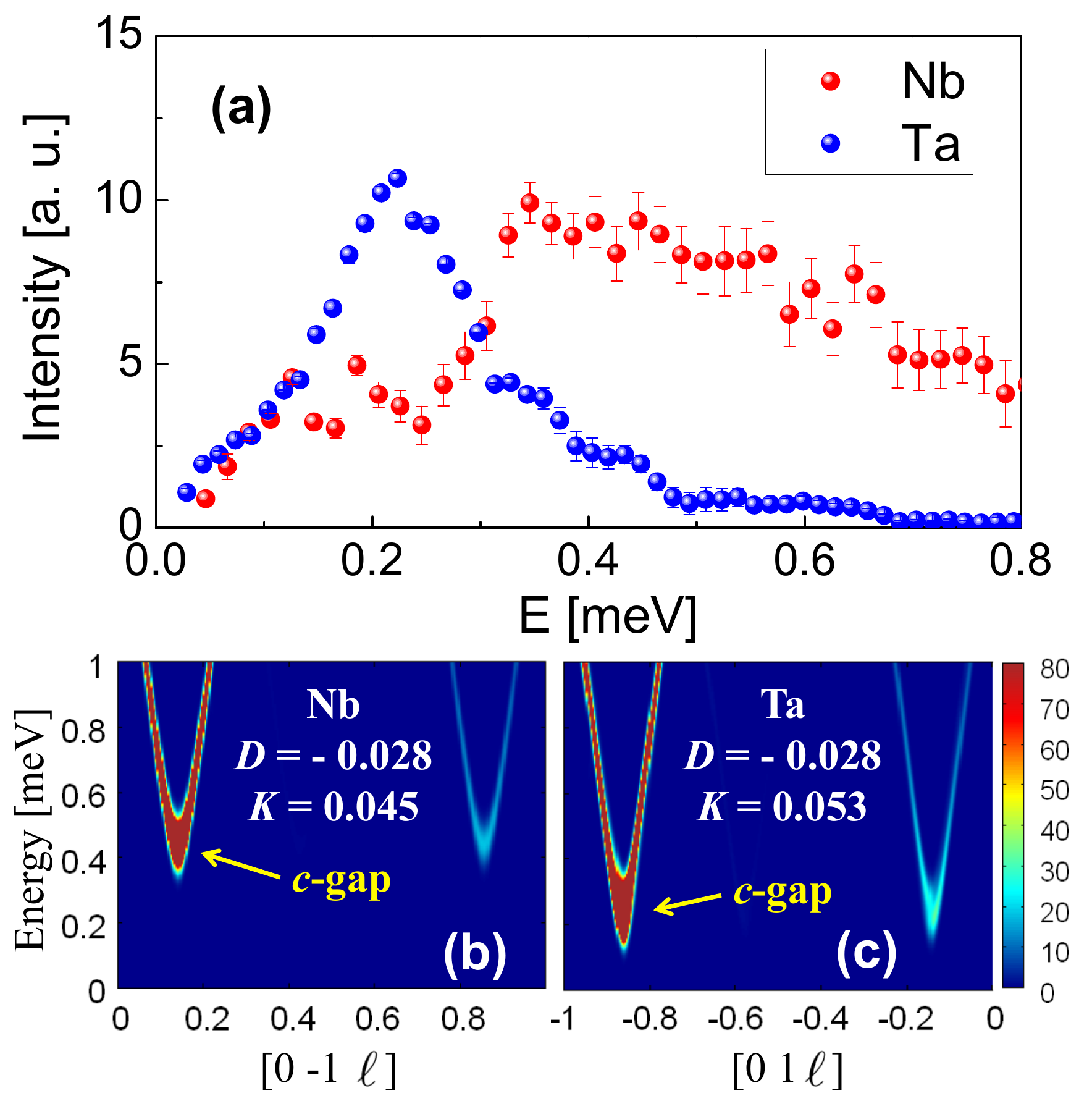}}
\caption{(Color online) (a)~Neutron scattering intensity versus the energy at the magnetic satellite position for Ta and Nb, obtained from 1D $Q$-cut (integration ranges $\Delta h=\Delta k=\Delta l=0.05$) measured on IN5 with a wavelength of 8 \AA. Below: Calculated scattering intensity of the $c$-branch only showing the $c$-gap of the Nb (b) and Ta (c) compounds using the revised model with the parameters of the caption of Fig. \ref{fig2}.}
\label{fig3}
\end{figure}

A quantitive determination of the parameters of the revised Hamiltonian can be achieved for the Nb compound. The intensity of the the first-order (0, k, $\ell \pm \tau$) and extra (0, k, $\ell \pm 3\tau$) satellites (Figs.~\ref{fig1} (d)) increases with the single-ion anisotropy and agrees with the experiments (see Fig. \ref{fig1}) for $|K|$=0.045$\pm$0.05 meV. This term is also at the origin of the anti-crossings observed in the spin-waves (at 2.3 meV for the $ab$-branches and at 1.7 meV for the $c$-branches): these spectral weight extinctions occur at the crossing point of branches with the same symmetry, emerging from the $+\tau$ and $-\tau$ elastic magnetic satellites, which are mixed by the single ion anisotropy $KS_z^2$ \cite{COQ78,SM}. The anti-crossings are more pronounced in the Ta case (see Fig. \ref{fig2} (b)) which qualitatively agrees with a higher value of $K$ \cite{SM}. Another signature of the presence of tiny anisotropies is a gap in the $c$-branch, which reaches its energy minimum at $\approx$ 0.4 meV for Nb \cite{STO11} and $\approx$ 0.2 meV for Ta as shown in Fig. \ref{fig3} (a). The variation of this gap is actually opposite for the easy-plane ($K>0$) and for the easy-axis ($K<0$) anisotropies. In the latter case, the $c$-gap increases with increasing $|K|$ and is already around 1 meV for $K$=0.045 meV, which is larger than the values observed experimentally. These values are actually compatible with an easy-plane anisotropy combined with a DM term (with $D<0$): the $c$-gap increases when $|D|$ increases and decreases when $K$ increases. The Nb $c$-gap is reproduced for $D$=-0.028 meV and $K$=0.045 meV (see Fig. \ref{fig3}(b)). The smaller gap of the Ta compound is in turn compatible with a larger $K$ than in the Nb one (see Fig. \ref{fig3}(c)). This analysis yields the extended model given in the caption of Fig. \ref{fig2} with a set of parameters reproducing the spin-waves for the Nb and Ta compounds (see Figs. \ref{fig2}(c-d)). Moreover, in the Nb case these parameters are fully compatible with the magnetic structure factors and with the Electron Spin Resonance measurements \cite{ZOR11} that have been reanalyzed carefully \cite{SM}. In real space, this revised model globally corresponds to a deviation of the angle between adjacent spins along the helix axis away from the ideal $2\pi\tau = 51.4^{\circ}$ value: It gets smaller near the local easy-planes and larger far from them (see Fig. \ref{fig1c} (b)). Thus, we conclude that the magnetic structure of the Fe langasites corresponds to a bunched helix, as reported in some rare-earth intermetallics \cite{JEN91}. 

Next, we analyze in more detail the first-order satellites along the [0 0 $\ell$]* line. These (0, 0, $\ell\pm \tau$) satellites are clearly visible in our data (see Fig.~\ref{fig1}), where they display the same temperature dependence as the (0, 1, $\ell \pm \tau$) magnetic ones \cite{SM}. In fact, the longitudinal polarization analysis confirms their magnetic origin (see Fig.~\ref{fig1}(c)). The observation of these satellites is very surprising since, in the case of a perfect 120$^{\circ}$ spin arrangement within the Fe triangles, there should be an extinction of the magnetic structure factor along the [0 0 $\ell$]* line. We note, that the forbidden (0, 0, $\tau$) satellite has also been observed by magnetic resonant X-ray diffraction (together with $2\tau$ and $3\tau$ satellites in the first Brillouin zone) \cite{SCA13}. This was attributed to a modulated out-of plane (butterfly) component of the magnetization produced by the in-plane intra-triangle DM vector. However, neutrons are blind to this component along [0 0 $\ell$]$^*$ since they are only sensitive to the magnetic components perpendicular to the scattering vector $Q$. Thus, the observation of these (0, 0, $\ell\pm \tau$) satellites in our neutron experiments clearly demonstrates that the relative orientation of the magnetic moments within the triangles deviates from 120$^{\circ}$, hence suggesting that the intra-triangle $J_1$ interactions are not equivalent. Our neutron data cannot  single out whether the distortion of this triangular arrangement is local (helically modulated from planes to planes) or global. However, only the latter case, implying a loss of the 3-fold symmetry axis of the $P321$ structure, is compatible with a macroscopic polarization measured along the two-fold axis. The loss of the 3-fold axis was already invoked from M\"ossbauer experiments and x-ray magnetic diffraction \cite{LYU11,SCA13}, and was an important ingredient for explaining the magnetoelectric excitations revealed by means of THz spectroscopy \cite{CHA13}. In addition, it is compatible with the phonon spectrum \cite{TOU15}. This finding has strong implications on the multiferroic properties of these systems, as we discuss in the following.

The P321 space group of the Fe langasites allows the magnetoelectric coupling of the form: 
$$F_{ME} =−\alpha\{P_x(\mathbf{S}_1-\mathbf{S}_2)\cdot \mathbf{S}_3-{1\over \sqrt{3}} P_y [(\mathbf{S}_1+\mathbf{S}_2)\cdot \mathbf{S}_3-2\mathbf{S}_1\cdot \mathbf{S}_2]\}. $$
where $y$ and $x$ stand for the $a$-axis and its perpendicular direction in the $ab$-plane, and $\mathbf{S}_1$ to $\mathbf{S}_3$ are the three spins on a triangle (see Fig. \ref{fig1c}). This coupling can have a purely electronic origin as discussed in \cite{BUL08} or emerge from magnetostriction effects in the symmetric Heisenberg exchange \cite{KHO15}. In any case, at the phenomenological level, its form is identical to the one discussed in \cite{BUL08,KHO10,KHO15}. The presence of such a coupling means that a spin-induced electric polarization has to be expected whenever the spin arrangement deviates from the perfect 120$^\circ$ structure \cite{footnote}. This polarization will lie in the $ab$-plane. Thus, the new magnetic structure deduced from our experiments is compatible with the polarisation measurements of Lee {\it et al.} \cite{LEE14} rather than those of Zhou {\it et al.} \cite{ZHO09}. This is also in tune with the DFT-based calculations reported in \cite{LEE10}, in which a magnetically-induced polarization was also obtained in the $ab$-plane due to the reduction of symmetry of the magnetic structure to C2'. In this work, however, this reduction was the result of spin-orbit coupling (likely via symmetric anisotropic exchange), and the precise magnetoelectric coupling behind such an electric polarization was not identified.

The enhancement of the electric polarization in the $ab$-plane by the application of a magnetic field has also been reported in \cite{LEE14}. This can be understood on the basis of the above coupling, although its connection with the revised magnetic structure is more subtle. 
In this case, the 120$^\circ$ spin-structure of each triangle is additionally deformed by the Zeeman coupling, which induces a (weak) ferromagnetic component and thus an extra local contribution to the electric polarization. In the perfect helical structure, however, these individual field-induced electric polarizations coming from the different layers of the helix rotate along the $c$ axis in such a way that they cancel each other. In the bunched helical structure, in contrast, this rotation is modified so that the cancelation is avoided, which eventually enables an overall finite polarization induced by the magnetic field. This will happen even if the initial (zero-field) structure corresponds to a perfect 120$^\circ$ arrangement within the Fe triangles. Thus, the magnetic-field induced electric polarization essentially relies on the bunching of the helix along the $c$-axis.

By carrying out neutron scattering experiments, we have identified subtle features in the magnetic order of Fe langasites, with drastic consequences on their properties. The first feature is a bunching of the helical modulation related to single-ion magnetocrystalline anisotropy, and the second one is the distortion of the 120$^{\circ}$ spin structure. At the phenomenological level, we have established that the magnetically-induced electric polarization at zero and finite magnetic fields results from a generic magnetoelectric coupling that is equivalent to that put forward by Bulaevskii {\it et al.} Our study thus sets the basis for understanding the multiferroic properties of Fe langasites representing model-case magnetically-induced ferroelectrics in triangle-based frustrated magnets. Finally, the close relation between the multifaceted chiral properties of the Fe langasites and their multiferroic potential is expected to motivate further studies.


\begin{acknowledgments}
We would like to acknowledge A. Hadj-Azzem and J. Debray for their help in the preparation of the samples, and P. Bordet, S. Grenier and O. C\'epas for fruitful discussions. This work was financially supported by Grant No. ANR-13-BS04-0013-01. 
\end{acknowledgments}

\end{document}


\title{Supplemental Material: Helical bunching and symmetry lowering inducing multiferroicity in Fe langasites}

\author{L. Chaix}
\affiliation{Institut Laue-Langevin, 6 rue Jules Horowitz, 38042 Grenoble, France}
\affiliation{Institut N\'eel, CNRS, 38042 Grenoble, France}
\affiliation{Universit\'e Grenoble Alpes, 38042 Grenoble, France}
\affiliation{Stanford Institute for Materials and Energy Sciences, SLAC National Accelerator Laboratory, Menlo Park, California 94025, USA}
\author{R. Ballou}
\affiliation{Institut N\'eel, CNRS, 38042 Grenoble, France}
\affiliation{Universit\'e Grenoble Alpes, 38042 Grenoble, France}
\author{A. Cano}
\affiliation{CNRS, Univ. Bordeaux, ICMCB, UPR 9048, F-33600 Pessac, France}
\author{S. Petit}
\affiliation{CEA, Centre de Saclay, /DSM/IRAMIS/ Laboratoire L\'eon Brillouin, 91191 Gif-sur-Yvette,France}
\author{S. de Brion}
\affiliation{Institut N\'eel, CNRS, 38042 Grenoble, France}
\affiliation{Universit\'e Grenoble Alpes, 38042 Grenoble, France}
\author{J. Ollivier}
\affiliation{Institut Laue-Langevin, 6 rue Jules Horowitz, 38042 Grenoble, France}
\author{L.-P. Regnault}
\affiliation{SPSMS-MDN, INAC, 38054 Grenoble, France}
\affiliation{Universit\'e Grenoble Alpes, 38042 Grenoble, France}
\author{E. Ressouche}
\affiliation{SPSMS-MDN, INAC, 38054 Grenoble, France}
\affiliation{Universit\'e Grenoble Alpes, 38042 Grenoble, France}
\author{E. Constable}
\affiliation{Institut N\'eel, CNRS, 38042 Grenoble, France}
\affiliation{Universit\'e Grenoble Alpes, 38042 Grenoble, France}
\author{C. V. Colin}
\affiliation{Institut N\'eel, CNRS, 38042 Grenoble, France}
\affiliation{Universit\'e Grenoble Alpes, 38042 Grenoble, France}
\author{A. Zorko}
\affiliation{Jo\v{z}ef Stefan Institute, Jamova 39, 1000 Ljubljana, Slovenia}
\author{V. Scagnoli}
\affiliation{Laboratory for Mesoscopic Systems, Department of Materials, ETH Zurich, 8093 Zurich, Switzerland}
\affiliation{Paul Scherrer Institute, 5232 Villigen PSI, Switzerland}
\author{J. Balay}
\affiliation{Institut N\'eel, CNRS, 38042 Grenoble, France}
\affiliation{Universit\'e Grenoble Alpes, 38042 Grenoble, France}
\author{P. Lejay}
\affiliation{Institut N\'eel, CNRS, 38042 Grenoble, France}
\affiliation{Universit\'e Grenoble Alpes, 38042 Grenoble, France}
\author{V. Simonet}
\affiliation{Institut N\'eel, CNRS, 38042 Grenoble, France}
\affiliation{Universit\'e Grenoble Alpes, 38042 Grenoble, France}

\maketitle
\onecolumngrid
\section{Validation of magnetic anisotropy terms by electron spin resonance}

Since our present neutron study has resulted in substantially different magnetic-anisotropy parameters of the investigated Fe langasites compared to our previous study that was based on the electron spin resonance (ESR)  \cite{ZorkoESR}, it is essential to check the compatibility between these two complementary techniques. The currently derived parameters are roughly an order of magnitude larger than the previous estimates. It should be stressed though that the previous ESR analysis only took into account both sources of magnetic anisotropy separately. As our current study has shown, the easy-plane single-ion anisotropy ($K>0$) effectively reduces the zero-field gap of the $c$-branch that is opened by the out-of-plane $c$-component of the DM interaction, $D$, which thus explains the reduced anisotropy parameters proposed in Ref.~\onlinecite{ZorkoESR}.

\begin{figure}
\resizebox{11cm}{!}{\includegraphics{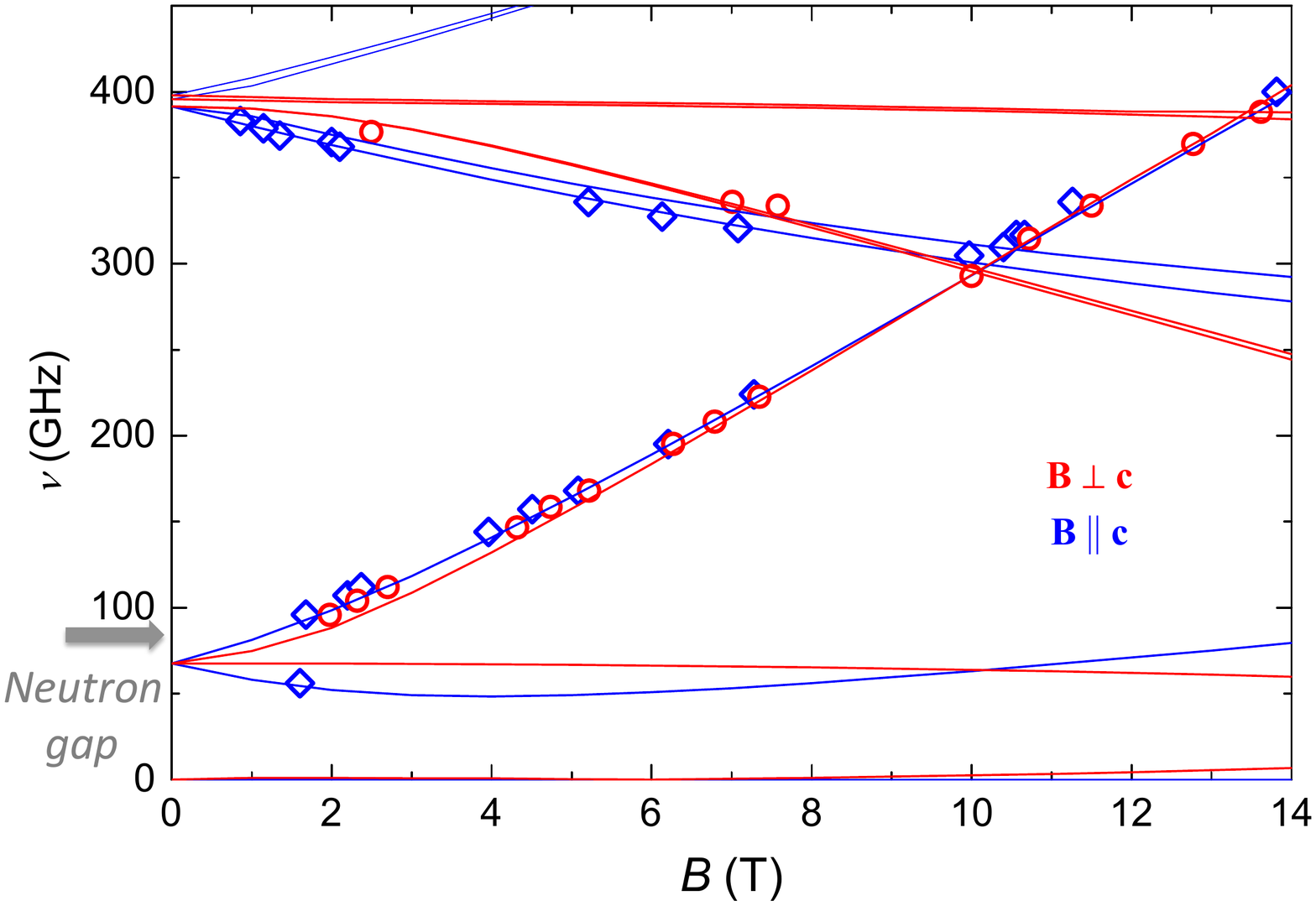}}
\caption{ The frequency-field diagram of measured (symbols) and simulated (lines) AFMR modes in Ba$_3$NbFe$_3$So$_2$O$_{14}$ at 4~K. The simulations correspond to the $c$-component DM anisotropy constants $D=-0.028$~meV and the easy-plane single-ion anisotropy $K=0.045$~meV. The gap of the $c$-branch determined by neutron measurements is indicated.}
\label{ESR}
\end{figure}

We have repeated the calculation of the AFMR modes detected in the previous ESR experiment \cite{ZorkoESR}. As can be seen in Fig.\ref{ESR}, with no adjustable parameters an excellent agreement is found for the values $D=-0.028$~meV and $K=0.045$~meV, which convincingly validates the results of our neutron scattering study.

We stress that the in-plane $ab$-component of the DM interaction, $D_{ab}$, does not affect the lowest AFMR modes and, therefore, can not be determined from the AFMR experiment. However, it does affect the ESR line-width anisotropy in the paramagnetic phase, which was found to saturate at the value of $\Delta B_{ab}/\Delta B_c=1.27$ at high temperatures \cite{ZorkoESR}. The above minimal model of the magnetic anisotropy, employing only $D$ and $K$ yields $\Delta B_{ab}/\Delta B_c=0.75$, therefore an additional anisotropy term is needed to explain the experiment. The isotropic $a$ term of the single-ion anisotropy cannot boost this ratio above 1, while $|D_{ab}|=2.5|D|$ reproduces the experiment. The minimal model of the magnetic anisotropy in Ba$_3$NbFe$_3$Si$_2$O$_{14}$ that reproduces all the experiments performed so far thus includes $D=-0.028$~meV, $|D_{ab}|=0.07$~meV and $K=0.045$~meV.  

\section{Deviation from a 120$^{\circ}$ magnetic arrangement and other models}

\begin{figure}
\resizebox{7cm}{!}{\includegraphics{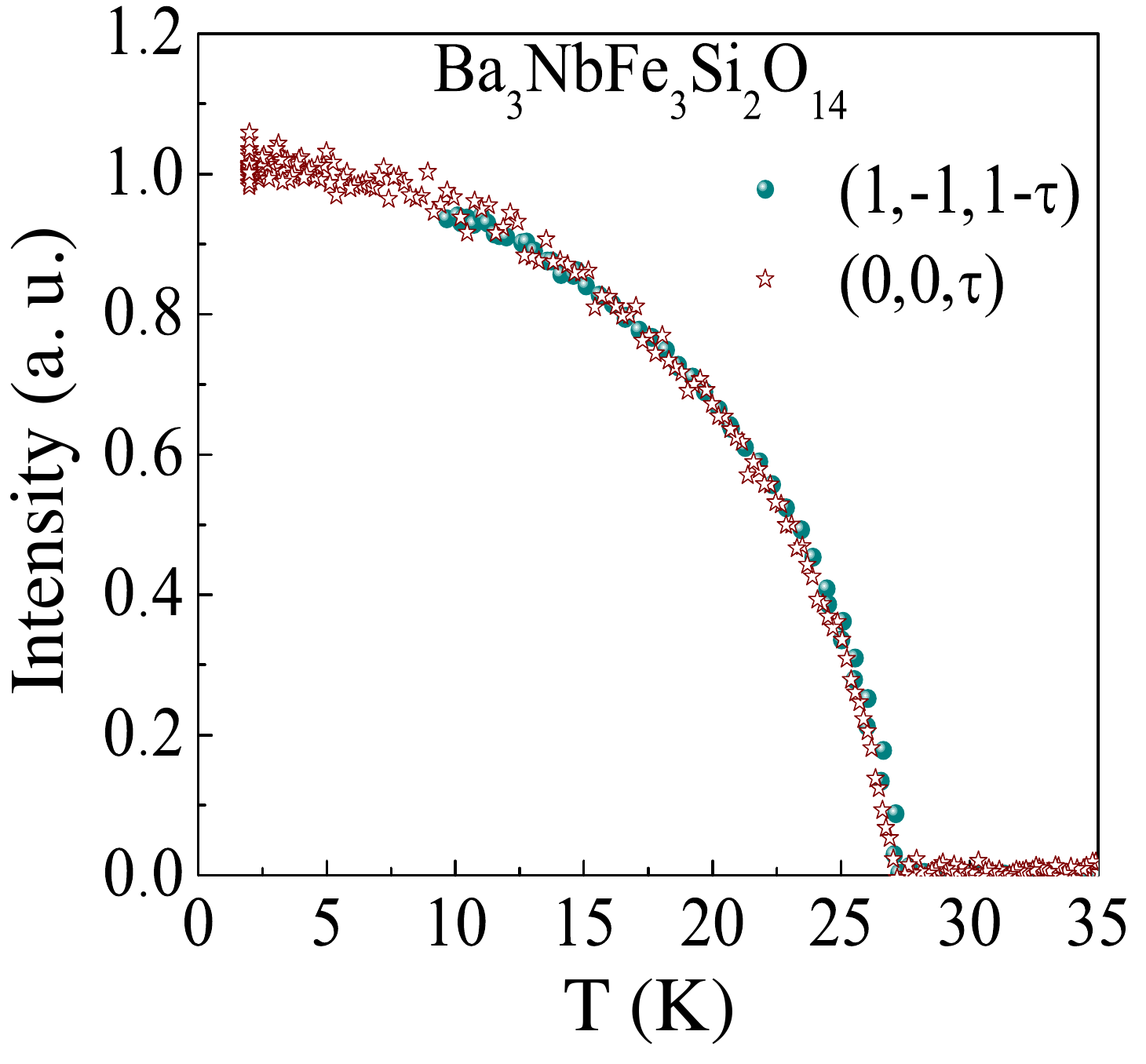}}
\caption{ Temperature dependence of the magnetic contributions of the (0, -1, 1-$\tau$) and (0, 0, $\tau$) satellites measured by neutron scattering and rescaled for comparison.}
\label{fig2SM}
\end{figure}

The magnetic structure of the Fe langasite is described by three magnetic helices propagating perpendicular to the ($a$, $b$) plane with a propagation vector (0, 0, $\tau$), based on magnetic moments at 120$^{\circ}$ from each other in the $ab$-plane. The magnetic structure factor of this structure, at the reciprocal space position $H=\pm\tau$ with $H$ a reciprocal lattice vector, writes :

$$\vec F_M(\vec{Q}=\vec{H}\pm\vec{\tau})=p\sum_{\nu=1,3}f(|\vec Q|)\vec m_{\pm \vec \tau,\nu}e{i\vec Q.\vec r_{\nu}}$$

with $p$=0.2696 10$^{-12}$ cm, $f$ the magnetic form factor of the Fe$^{3+}$ atoms, $\vec r_{\nu}$ the position of the atom $\nu$ in the cell: $\vec{r}_{1} = (x, 0, \frac{1}{2})$, $\vec{r}_{2} = (0, x, \frac{1}{2})$ and $\vec{r}_{3} = (-x, -x, \frac{1}{2})$ with $x$=0.2496. The Fourier components of the magnetization distribution for the atom $\nu$ is $m_{\pm \vec \tau,\nu}=[{\mu  \hat{u} \pm i \mu \hat{v}\over 2}] e^{\mp i \phi_{\nu}}$ with $\mu$ the magnetic moment amplitude, ($\hat u,\hat v$) and orthonormal basis in the $ab$-plane, and $\phi_{\nu}$ a phase which should account for the relative 120$^{\circ}$ dephasing between the three Bravais lattices.

For $H=(0, 0, \ell)$, the magnetic structure factor becomes zero: 
$$\vec F_M(0, 0, \ell\pm\tau)=pf(|\vec Q|)\sum_{\nu=1,3}[\frac{\mu \hat{u} \pm i \mu \hat{v}}{2}] e^{\mp i \Phi_{\nu}} \; e^{i 2\pi \frac{(\ell \pm \tau)}{2}} =pf(|\vec Q|)[\frac{\mu \hat{u} \pm i \mu \hat{v}}{2}] e^{i 2\pi \frac{(\ell \pm \tau)}{2}}(1 + e^{\mp i \frac{2\pi}{3}} + e^{\mp i \frac{4\pi}{3}})=0.$$ 

For in-plane magnetic moments, the only way to observed a magnetic signal on the (0, 0, $\ell\pm\tau$) satellites, implies a deviation from the 120$^{\circ}$ arrangement of the magnetic moments. The measured temperature dependence of the (0, 0, $\tau$) extra-satellites, produced by the deformation of the 120$^{\circ}$ magnetic arrangement is shown in Fig. \ref{fig2SM}. It follows the temperature dependence of strong first order magnetic satellites.

For completeness, we have tried two other models: the first one involves, in addition to the Hamiltonian of Eq.1, a biquadratic term of the form $[(S_1.S_2)(S_2 - S_3).S_1 + S_2.S_3(S_3 - S_1).S_2 + (S_3.S_1)(S_1 - S_2).S_3]$ due to the spin-phonon coupling, (ii) the second model, called KSEA (for Kaplan-Shekhtman-Entin-Wohlman-Aharony), takes into account terms quadratic in $D$, which leads to experimental observations such as 3$\tau$ satellites and spectral weight extinction in the spin-waves \cite{SHE93,ZHE99}. We found that none of these models quantitatively reproduce our data, therefore we discarded them.

\section{Note on the chirality}

The excellent agreement between the calculated and observed spectral weight extinction in the spin-waves when including a single-ion anisotropy in the Hamiltonian is only observed for one triangular chirality (anticlockwise 120 $^{\circ}$ rotation of the spins on each triangle), in agreement with previous determination of the magnetic chirality \cite{SIM12}. Both Ta and Nb single-crystals thus have the same triangular chirality. Their structural chirality is opposite, as determined from anomalous X-ray scattering. This structural chirality, in turn, imposes the helical chirality to be opposite in both compounds.

\section{Neutron polarimetry}

The simplest neutron polarimetry technique to probe the chiral scattering was first introduced by Moon, Riste and Koehler \cite{Moon1969}, and is called longitudinal polarization analysis (LPA). In this case the final and initial polarizations are parallel, which can be achieved typically on a triple-axis spectrometer with polarizing monochromator/analyser ({\it e.g.} Heusler crystals). In addition, two flippers select the polarization states $+$ and $-$ (parallel or antiparallel to the polarization axis). The spin-flip terms (scattering processes changing the sign of the polarization), 
$$({d^2\sigma\over d\Omega dE_f})^{+-}=\sigma^{+-} \mathrm{~and~}({d^2\sigma\over d\Omega dE_f})^{-+}=\sigma^{-+}$$ 
and the non-spin-flip terms (scattering processes leaving the sign of the polarization unchanged), 
$$({d^2\sigma\over d\Omega dE_f})^{++}=\sigma^{++} \mathrm{~and~} ({d^2\sigma\over d\Omega dE_f})^{--}=\sigma^{--}$$ 
of the partial differential cross-section can be measured independently. 

\vskip 0.2 cm

In neutron polarimetry, a right-handed coordinated system is usually chosen with the $x-$axis along the scattering vector $\vec Q$, the $y-$axis in the scattering plane and the $z-$axis perpendicular to the scattering plane, so that the magnetic interaction vector has zero $x$ component. In the following, we shall use the simplified notations:
the nuclear cross-section $\sigma_N$ is proportional to $|N|^2$ with $N$ the nuclear structure factor. The magnetic cross-section $\sigma_M$ is proportional to $|M_{\perp}|^2$ with $M_{\perp}$ the magnetic interaction vector, which is the projection of the magnetic structure factor $F_M$ onto the plane perpendicular to the scattering vector $Q$. The chiral cross-section $\sigma_{ch}$ is proportional to $M_{ch}$ defined as:

$$M_{ch}=i(M_{\perp}^{Z*}M_{\perp}^{Y}-M_{\perp}^{Y*}M_{\perp}^{Z}).$$

These contributions can then be easily determined from linear combinations of the spin-flip and non-spin-flip cross-sections, knowing the initial polarisation $P_i$ :

$$|N|^2\propto\sigma_x^{++}=\sigma_x^{--}$$

$$|M_{\perp}|^2\propto{\sigma_x^{+-}+\sigma_x^{-+}\over 2}$$

$$M_{ch}\propto{\sigma_x^{+-}-\sigma_x^{-+}\over 2P_i}.$$

Alternatively, one can use an initial unpolarized beam (produced by a graphite monochromator for instance) and perform polarization analysis. In this case, the chiral scattering can be derived from the difference:
$${\sigma_x^{0-}-\sigma_x^{0+}\over 2}=M_{ch}$$
where the symbol 0 is meant for zero beam polarization of the incoming neutrons.

The flipping ratio of R $\approx$ 16 was determined experimentally by measuring the neutron counts in the Spin-Flip (SF) and Non Spin-Flip (NSF) channels on different nuclear Bragg peaks. This value was taken into account in the IN22 data analysis to correct the intensities measured in the SF and NSF channels from the imperfectly polarized neutron beam. 

\section{Details on the Neutron Data Analysis}

The analysis of the neutron data was performed using the {\it spinwave} software developped at LLB by S. Petit. The starting point is a bilinear spin Hamiltonian
\begin{eqnarray*}
{\cal H} &=& 
{\cal H}_{\mbox{CEF}} + \frac{1}{2} \sum_{i,j} \vec{S}_i {\cal J}_{i,j}  \vec{S}_j 
\end{eqnarray*}
where $ {\cal J}_{i,j}$ denotes the effective exchange coupling (including standard exchange, anisotropic, Dzyaloshinskiy-Moryia or bond directed coupling); ${\cal H}_{\mbox{CEF}}$ is a single ion anisotropy term :
\begin{equation*}
{\cal H}_{\mbox{CEF}} = 3B_{20} \left( S_z^2 - S(S+1) \right).
\end{equation*}
At the mean field level : 
\begin{eqnarray*}
{\cal H}_{\rm MF} &=& \sum_i {\cal H}_i \\
{\cal H}_i  &=&
{\cal H}_{\mbox{CEF}} +\vec{S}_i \sum_{j} {\cal J}_{i,j} \langle \vec{S}_j \rangle. 
\end{eqnarray*}
Starting from a random configuration for the observables $\langle \vec{S}_i\rangle$, the contribution to ${\cal H}_{\rm MF}$ at site $i$ (in the unit cell) is diagonalized in the Hilbert space defined by the $\left\{ | S_z \rangle \right\}, S_z=-S,...,S$ basis vectors ($2S+1$ vectors). This yields the energies $E_{i,n}$ and the wave functions $\vert \phi_{i,n} \rangle$. The updated expectation values, $\langle \vec{S}_i \rangle'$, at each step of the iteration procedure, are given by: 
\begin{equation*}
\langle \vec{S}_i  \rangle' =  \sum_{n} \frac{e^{-E_{i,n}/k_B T}}{Z} \langle  \phi_{i,n}  | \vec{S}_i | \phi_{i,n} \rangle
\end{equation*}
with
\begin{equation*}
Z =\sum_{n} \exp\left({-E_{i,n}/k_B T}\right).
\end{equation*}
These are then used to proceed to site $j$, and this is repeated until convergence. 

In the present case, the starting point is a spin arrangement with a magnetic cell 7 times larger than the nuclear one, which is a correct description for the Nb langasite and only approximate for the Ta langasite (propagation vectors respectively equal to $\approx$1/7 and $\approx$1/7.22). This allowed us to find that, without single-ion anisotropy, the mean field procedure converges to a perfect helix where the moments are at 120$^{\circ}$ from each other. Including the single ion anisotropy we find that the helix becomes bunched (still with a 120$^{\circ}$ magnetic order within the triangles). Then, imposing the loss of the 3-fold axis (by taking different $J_1$ on one triangle), we find that the 120$^{\circ}$ arrangement is perturbed. 
After this mean field step, the same software computes the neutron magnetic structure factors for all the satellites that can be compared with the measured ones. Last the spin-waves spectrum (based on the Holstein-Primakov description) of the minimized spin configuration is obtained and compared to the experiment. We found an excellent agreement between the spin configuration imposed by our model Hamiltonian, the extra satellites and the observed extinction in the spin-waves. The single-ion anisotropy produces a bunching of the helix, the rise of 3$\tau$ satellites, and the extinction in the spin-waves. The loss of the 3-fold axis additionally produces a deformation of the 120$^{\circ}$ arrangement and the rise of the first order satellites along the [0 0 $\ell$]$^*$ direction.

The biquadratic coupling 
\begin{equation*}
V = \sum_{ijkl} \lambda_{ijkl}~\vec{S}_i \cdot \vec{S}_j ~\vec{S}_k \cdot \vec{S}_l
\end{equation*} 
can be taken into account in the software. Assuming the symmetry properties:
\begin{eqnarray*}
\lambda_{ijkl} & = & \lambda_{jikl} \\
\lambda_{ijkl} & = & \lambda_{ijlk} \\
\lambda_{ijkl} & = & \lambda_{klij} 
\end{eqnarray*}
the non-linear nature of the biquadratic coupling is treated by a decoupling scheme, ending up with bilinear couplings only. The following approximation is performed: 
\begin{equation*}
\left\{
\begin{array}{lll}
V &\approx &\sum_{ij}\sum_{ab} S_i^a \left( 2 \sum_{kl} \lambda_{ijkl}~ \vec{\eta}_k ~\vec{\eta}_l ~\delta_{ab} ~+ ~4 \left(\sum_{kl} \lambda_{ikjl}~ \eta^a_k~ \eta^b_ l \right) \right) S_j^b \\
\vec{\eta}_i & = &\langle \vec{S}_i \rangle
\end{array}
\right.
\end{equation*}
leading, from the formal point of view, to a new effective anisotropic exchange term. 
